\documentclass[prl,
 reprint,
superscriptaddress,
 amsmath,amssymb,
 aps,
]{revtex4-2}

\usepackage{graphicx}
\usepackage{dcolumn}%
\usepackage{bm}%
\usepackage[dvipsnames]{xcolor}
\usepackage{soul}
\usepackage{afterpage}
\usepackage{xr-hyper}
\usepackage{hyperref}
\usepackage{amsmath}




\begin{document}

\title{Kohler's Rule Holds in the Strange Metallic Hubbard Model}%

\author{Rebekah M. Jin}
\email{rebjin@stanford.edu}
\affiliation{Department of Applied Physics, Stanford University, CA 94305, USA}
\affiliation{Stanford Institute for Materials and Energy Sciences,
SLAC National Accelerator Laboratory, 2575 Sand Hill Road, Menlo Park, CA 94025, USA}

\author{Emily Z. Zhang}
\affiliation{Stanford Institute for Materials and Energy Sciences,
SLAC National Accelerator Laboratory, 2575 Sand Hill Road, Menlo Park, CA 94025, USA}
\affiliation{Geballe Laboratory for Advanced Materials, Stanford University, California 94305, USA}

\author{Jixun K. Ding}
\affiliation{School of Physics and Astronomy, Monash University, Clayton, VIC 3150, Australia}
\affiliation{Department of Applied Physics, Stanford University, CA 94305, USA}
\affiliation{Stanford Institute for Materials and Energy Sciences,
SLAC National Accelerator Laboratory, 2575 Sand Hill Road, Menlo Park, CA 94025, USA}

\author{Brian Moritz}
\affiliation{Stanford Institute for Materials and Energy Sciences,
SLAC National Accelerator Laboratory, 2575 Sand Hill Road, Menlo Park, CA 94025, USA}

\author{Thomas P. Devereaux}
\email{tpd@stanford.edu}
\affiliation{Stanford Institute for Materials and Energy Sciences,
SLAC National Accelerator Laboratory, 2575 Sand Hill Road, Menlo Park, CA 94025, USA}
\affiliation{
Department of Materials Science and Engineering, Stanford University, Stanford, CA 94305, USA}
\affiliation{Geballe Laboratory for Advanced Materials, Stanford University, California 94305, USA}

\date{\today}%

\begin{abstract} 
Unconventional magnetotransport violating Kohler's rule has been reported in many strongly correlated strange metals, but the exact origin of these violations is yet unknown.
To investigate this question, we use determinant quantum Monte Carlo to compute the magnetoresistance of the hole-doped Hubbard-Hofstadter model. Despite $T$-linear resistivity exceeding the Mott-Ioffe-Regel limit and the nominal absence of well-defined quasiparticles, we find Kohler’s rule unexpectedly holds across a wide range of on-site interaction strengths and hole dopings.
In particular, Kohler emerges below a crossover temperature $T^\star \approx t$ identified from the double occupancy and specific heat, coinciding with the onset of local singlet correlations between lattice sites.
We thus propose that Kohler scaling in this regime reflects a more general form of many-body quantum coherence. Our results show that strong correlations and strange metallicity do not themselves generically produce anomalous magnetoresistance.

\end{abstract}

\maketitle

\paragraph{Introduction.---} Strongly correlated electron systems are one of a few longstanding challenges in modern condensed matter physics. 
Defying understanding by conventional single-particle theories,
many strongly correlated materials exhibit anomalous electrical transport properties: $T$-linear resistivity exceeding the Mott-Ioffe-Regel (MIR) limit \cite{gurvitch_resistivity_1987,martin_normal-state_1990,takagi_systematic_1992} associated with Planckian dissipation \cite{Bruin804,legros_universal_2019}, strongly temperature- and doping-dependent Hall conductivity \cite{Hall1,Hall2,Hall3}, and scale invariant linear magnetoresistance (MR) \cite{giraldo-gallo_scale-invariant_2018}. These results imply that the collective, charge-carrying excitations in such materials are not well-described by the conventional quasiparticles of Fermi liquid theory.

The absence of a comprehensive theory of transport invites the thorough examination of empirical scaling laws, each a valuable macroscopic probe of otherwise intractable microscopic scattering mechanisms.
A powerful measurement is the magnetoresistance 
\begin{equation}
\text{MR}(B) = \frac{\rho(B) - \rho_0}{\rho_0},
\label{eq:MR}
\end{equation}
the relative change in the resistivity $\rho$ from its zero-field value $\rho_0$ induced by an applied magnetic field $B$. Semiclassically, MR may be thought of as arising from circulating particles; charge carriers are localized by the Lorentz force into cyclotron orbits, 
generating a positive MR sensitive to both the shape of the Fermi surface and the nature of the orbiting particles.

This cyclotron picture is codified in a semiclassical Boltzmann transport-derived scaling law known as Kohler's rule: at low magnetic fields, the MR$=F(B/\rho_0)$, where $F$ is a temperature-independent function set solely by the electronic structure and geometry of the material \cite{kohler1938, luo_kohlers_2001}. Equivalently, the field dependence of the MR enters only through the parameter $B/\rho_0$, a proxy for the average orbital angular deflection between particle collisions. In a material obeying Kohler's rule, MR curves plotted against $B/\rho_0$ for different temperatures all overlap, or ``collapse'' \cite{pippard1989}.

Kohler scaling is well-established in typical metals \cite{ziman1960, pippard1989}. In the strongly interacting regime, however, where semiclassical assumptions often lose their validity, the rule need not be obeyed. In the high-$T_c$ superconducting cuprates, for example,
violations have been reported in underdoped YBCO and LSCO \cite{harris_violation_1995, kimura_LSCO_1996, ando_magnetoresistance_2002} as well as overdoped Bi2201 and Tl2201 \cite{kohler_Bi2201_2021}, while Kohler's rule is restored in overdoped, non-superconducting LSCO \cite{kimura_LSCO_1996} and in the pseudogap phase of underdoped Hg1201 \cite{kohlerHg1201}. Beyond the cuprates, deviations from or modifications to Kohler's rule have been observed in topological semimetals and heavy Fermion superconductors as well \cite{xu_extended_2021, nakajima_non-fermi_2007}.

Anomalous MR scaling is commonly interpreted as evidence of multiple carrier types with distinct scattering rates \cite{anderson_hall_1991, coleman_phenomenological_1996, harris_violation_1995, mckenzie_violation_1998, ando_magnetoresistance_2002}, anisotropic scattering in temperature-dependent ``hot spots" across the Fermi surface \cite{hotspot1, hotspot2, hotspot3}, or a breakdown of the quasiparticle picture altogether \cite{nakajima_non-fermi_2007, hussey_phenomenology_2008}. 
Still, whether Kohler violation is a direct reflection of strongly correlated physics remains an open question.

In this paper, we study the Hubbard-Hofstadter model, a minimal lattice Hamiltonian of interacting electrons in a magnetic field. 
The Hubbard model and its extensions have been instrumental in disentangling the specific contribution of strong correlations and reproduce key cuprate phenomenology \cite{huang_stripe_2018, huang_strange_2019, intrinsic_2025}.
Using determinant quantum Monte Carlo (DQMC), an unbiased and numerically exact algorithm, we simulate the model at varying on-site Coulomb interaction strengths, magnetic field strengths, and hole dopings, a regime largely unexplored relative to the half-filled case. 
Below a cutoff magnetic field $B^\star$, identified from the transverse conductivity $\sigma_{xy}$, and a coherence temperature $T^\star$, determined from the double occupancy and specific heat, Kohler’s rule is surprisingly robust across our parameter space. It survives even where quasiparticles are nominally ill-defined and the resistivity substantially exceeds the MIR limit.
We argue this robustness reflects a more general notion of many-body quantum coherence rather than orbiting quasiparticles, an alternative explanation of Kohler scaling that does not depend on inapplicable semiclassical reasoning.

\textit{Methods.---} 
We study the single-band Hubbard-Hofstadter model
\begin{multline}
H = -t \sum_{ \langle ij\rangle  \sigma}  \exp\left[i\varphi_{ij}\right]c_{i \sigma}^\dagger c_{j \sigma} - \mu \sum_{i \sigma} c^{\dagger}_{i \sigma}c_{i\sigma} + \\
U\sum_{i}\left(c^\dagger_{i \uparrow}c_{i\uparrow} - 1/2 \right)\left(c^\dagger_{i \downarrow}c_{i\downarrow} - 1/2\right) \label{eq:hamiltonian}
\end{multline}
on a two dimensional square lattice, where $t$ is the hopping integral between nearest neighbor sites $i$  and $j$, $\mu$ is the chemical potential and $U$ is the on-site Coulomb interaction strength. $c_{i \sigma}^{\dagger}$/$c_{i \sigma}$ is the creation/annihilation operator for an electron on site $i$ with spin $\sigma$. 
The Peierls phase
\begin{equation}
\varphi_{ij} = \frac{2\pi}{\Phi_0}\int_{\mathbf{r_i}}^{\mathbf{r_j}} \textbf{A} \cdot d\bm{\ell},
\end{equation}
contributes a uniform out-of-plane orbital magnetic field $\mathbf{B}=B \hat{z}$, where the path integral is taken over the shortest straight line path between sites $i$ and $j$. For the magnetic vector potential $\mathbf{A}= -\alpha B y\hat{x} + (1-\alpha) B x\hat{y}$, we use the symmetric gauge $\alpha=1/2$.

In DQMC, we simulate this Hamiltonian on an $N_x=8$ by $N_y=8$ square lattice with lattice constant $a=1$ for interaction strengths $U/t=2, 4,$ and $6$ at particle density $\langle n\rangle = 0.8$. Stronger interactions simultaneously suppress the MR and worsen the fermion sign, capping the accessible range of $U$.
We also sweep a range of hole dopings $\langle n\rangle = 0.8, 0.85, 0.9, 0.95,$ and $0.97$ at $U/t=4$. We enforce modified periodic boundary conditions \cite{periodic_bcs}, and requiring a single-valued wave function yields the flux quantization condition 
$B = n_f \Phi_0/ N_x N_y a^2$ -- only an integer number $n_f$ of magnetic flux quanta $\Phi_0 = h/e$ may thread the entire system. 

DQMC measures the unequal imaginary time electric current-current correlator
\begin{equation}
\chi_{\alpha\beta}(\tau) = \frac{1}{V}\langle J_{\alpha}(\tau) J_{\beta}(0)\rangle
\end{equation}
where $\tau$ is imaginary time and $\alpha, \beta$ index spatial directions $x, y$. Kubo formulas relate these Matsubara time correlators to real frequency-resolved transport coefficients \cite{mahan_many_particle_2000}. For the longitudinal response $\alpha=\beta$, we use MaxEnt analytic continuation \cite{jarrell_bayesian_1996} to obtain the imaginary component of the retarded correlator $\chi_{xx}^R(\omega)$, directly proportional to the real part of the optical conductivity $\sigma_{xx}(\omega)$. For $\alpha \neq \beta$, however, the analytically continued quantity is instead proportional to the imaginary part of $\sigma_{xy}(\omega)$ and thus antisymmetric in frequency. As MaxEnt cannot directly extract non-positive definite spectra, we use the subtraction method, first continuing a composite $xx$ and $xy$ correlator before subtracting off the longitudinal component \cite{reymbaut_maximum_2015, reymbaut_maximum_2017}. 

The off-diagonal correlator, however, has a poor signal-to-noise ratio that worsens with lower temperature, causing standard MaxEnt to yield asymmetric spectra that fail to obey the physical odd parity. We therefore also apply a newly developed constrained MaxEnt method, which uses constrained convex optimization to enforce required symmetries.
Accounting for transverse transport is particularly important at large magnetic fields, which produce an off-diagonal conductivity comparable in magnitude to the longitudinal component. Detailed analytic continuation methods can be found in the Supplemental Material Sec. II and in an upcoming paper.

We finally Kramers-Kronig transform the constrained MaxEnt result to obtain $\mathrm{Re}[\sigma_{xy}]$
before taking the $\omega \rightarrow 0$ limit of our real conductivity spectra to obtain D.C. $\sigma_{xx}$ and $\sigma_{xy}$. The longitudinal resistivity follows as
\begin{equation}
    \rho_{xx} = \frac{\sigma_{xx}}{\sigma_{xx}^2 + \sigma_{xy}^2}
\end{equation}
and the magnetoresistance is calculated from Eq. \ref{eq:MR}. To reduce finite-size effects, we use magnetic field strength $B = 1/64$ in units of $\Phi_0/a^2$ for $\rho_0$ \cite{periodic_bcs}.

\begin{figure*}[!htbp]
    {\includegraphics[width=0.8\textwidth]{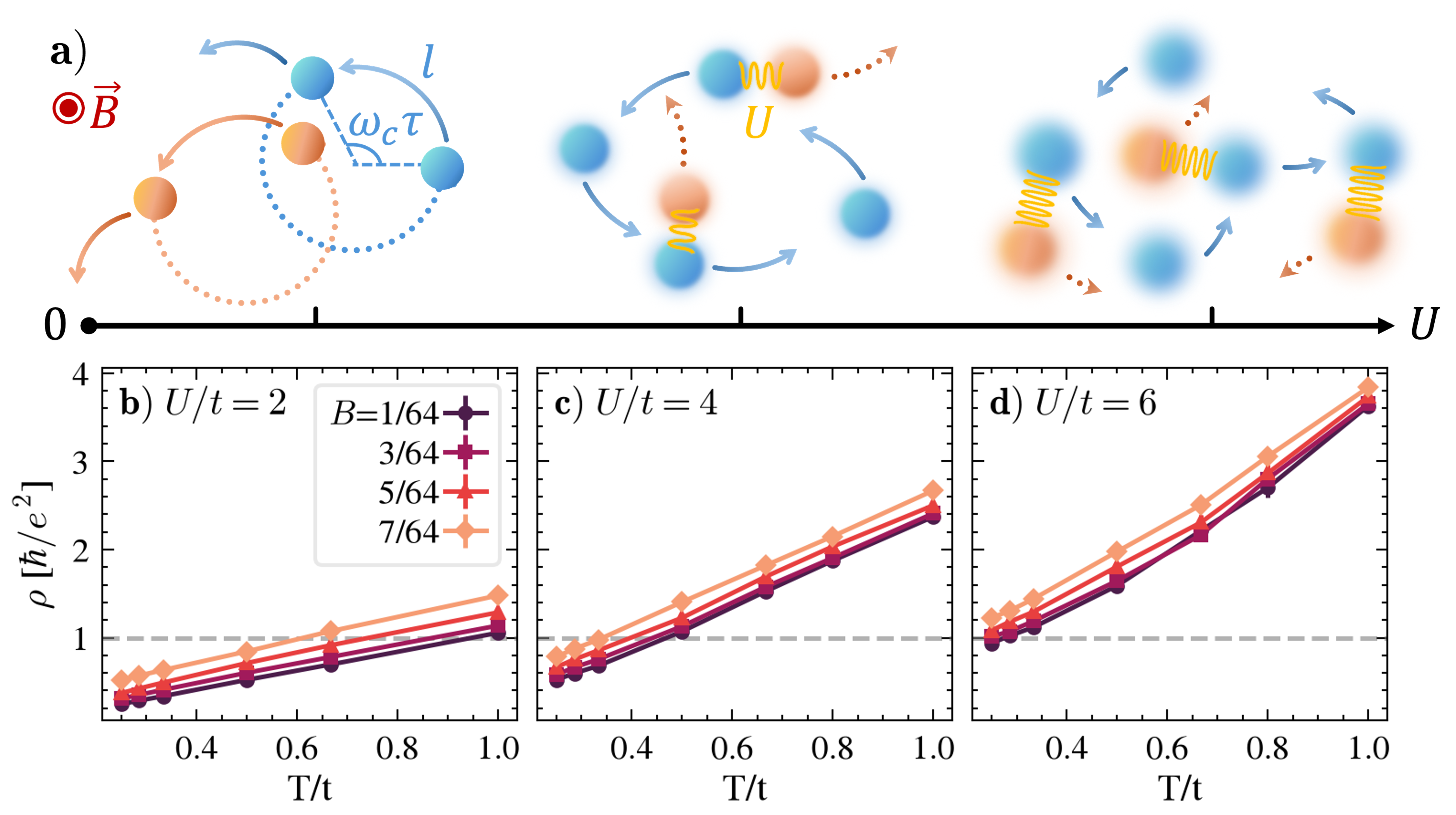}}
    \caption{Semiclassical magnetoresistance picture and Hubbard-Hofstadter resistivity versus interaction strength $U$. (a) Schematic illustration of semiclassical magnetoresistance in an interacting model as a function of $U$. At weak coupling (left), electrons undergo independent cyclotron orbits of mean free path $l$ with cyclotron frequency $\omega_c$ and scattering time $\tau$. Stronger coupling (center) shortens the mean free path, and at strongest $U$ (right) quasiparticles may be shorter-lived yet still traverse well-defined orbits governed by a single scattering time.
    (b--d) Temperature dependence of the D.C. longitudinal resistivity $\rho$ for $U/t = 2$, $4$, $6$, respectively. The particle density $\langle n \rangle = 0.8$ for all panels. The magnetic field $B$ is in units of $\Phi_0/a^2.$}
    \label{fig:rho}
\end{figure*}

\textit{Results.---} 
To establish a baseline picture of magnetoresistance in an interacting model, we begin by reviewing the semiclassical schematic in Fig. \ref{fig:rho}(a). At weak coupling, an out-of-plane magnetic field $B$ confines electrons into independent cyclotron orbits, each deflecting a mean angle $\omega_c \tau$ between scattering events. In Kohler's rule, this angle (or a proportional quantity $B/\rho_0$) 
sets the scale of a positive MR, which is generally quadratic in both $B$ and $T$ for conventional metals \cite{pippard1989}.
Provided that mass-renormalized quasiparticle wave packets 
survive into the interacting regime, strengthening $U$ should straightforwardly decrease $\tau$ and suppress the MR, while Kohler's rule remains unaffected.

We then show the D.C. longitudinal resistivity $\rho_{xx}(T)$ of the interacting Hubbard-Hofstadter model in Fig. \ref{fig:rho}(b-d) for $U/t=2-6$. As expected, both $U$ and $B$ shift the resistivity curves upward, so MR$(B)>0$. However, $\rho$ is not quadratic but robustly $T$-linear and exceeds the MIR limit $\hbar/e^2$,
hallmarks of strange metallic transport and the breakdown of the conventional quasiparticle picture of Fig. \ref{fig:rho}(a) \cite{phillips_stranger_2022}.
Whether Kohler's rule survives in this regime is therefore non-trivial.

\begin{figure}[!htbp]
    {\includegraphics[width=0.9 \columnwidth]{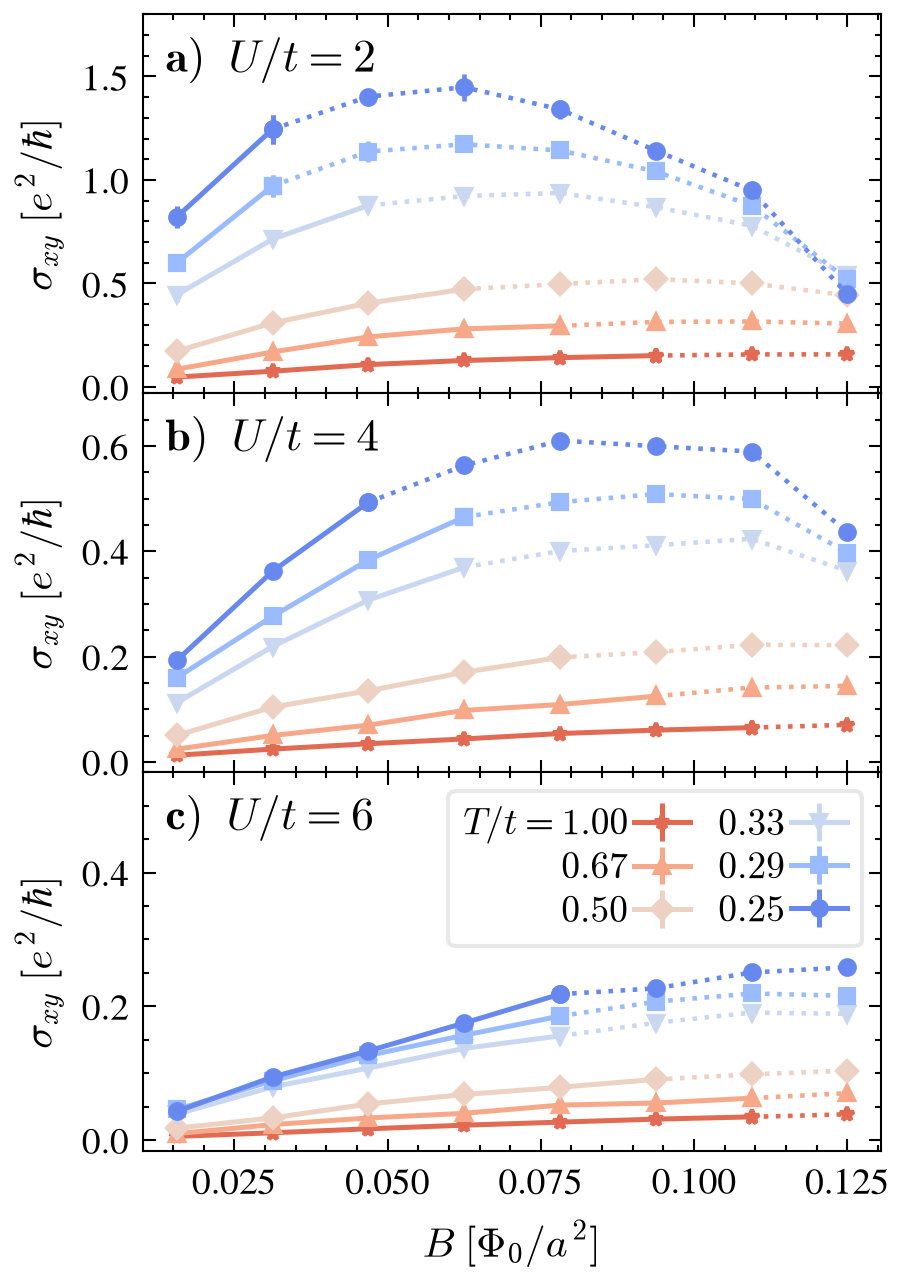}}
    \caption{Magnetic field and temperature dependence of the D.C. transverse conductivity $\sigma_{xy}.$ (a-c) are interaction strengths $U/t = 2, 4, 6$, and particle density $\langle n\rangle =0.8$ for all panels. $\sigma_{xy}$ is approximately $B$-linear below the cutoff field $B^\star(U, T, n)$. For each temperature, solid and dotted connecting lines indicate fields below and above $B^\star$, respectively.}
    \label{fig:sigma}
\end{figure}

As Kohler's rule is derived assuming low magnetic fields $\omega_c \tau \ll 1$ \cite{luo_kohlers_2001}, we determine a corresponding weak-field limit in our data. 
Using the D.C. transverse conductivities $\sigma_{xy}(B)$ shown in Fig. \ref{fig:sigma}, we define a maximum cutoff magnetic field $B^\star(U, T, n)$ below which $\sigma_{xy}$ is approximately $B$-linear, the leading order term allowed by Onsager-Casimir reciprocity \cite{onsager_reciprocal_1931, casimir_1945}.
We further observe that $\sigma_{xy}$ peaks above $B^\star$ at higher fields as the system becomes more resistive with greater $T$ and $U$. In the Drude model, this maximum occurs at $\omega_c \tau = 1$ or $B_{\mathrm{max}} \propto 1/\tau \propto \rho$ \cite{ashcroft_mermin}, consistent with the shifting peak positions in our data. 
This further suggests that $\omega_c \tau < 1$ below $B^\star$, matching the Kohler weak-field criterion. Full $\sigma_{xy}$ and cutoff fields are shown in the Supplemental Material Sec. V.

\begin{figure*}[t]
    \centering
    \includegraphics[width=0.9\textwidth]{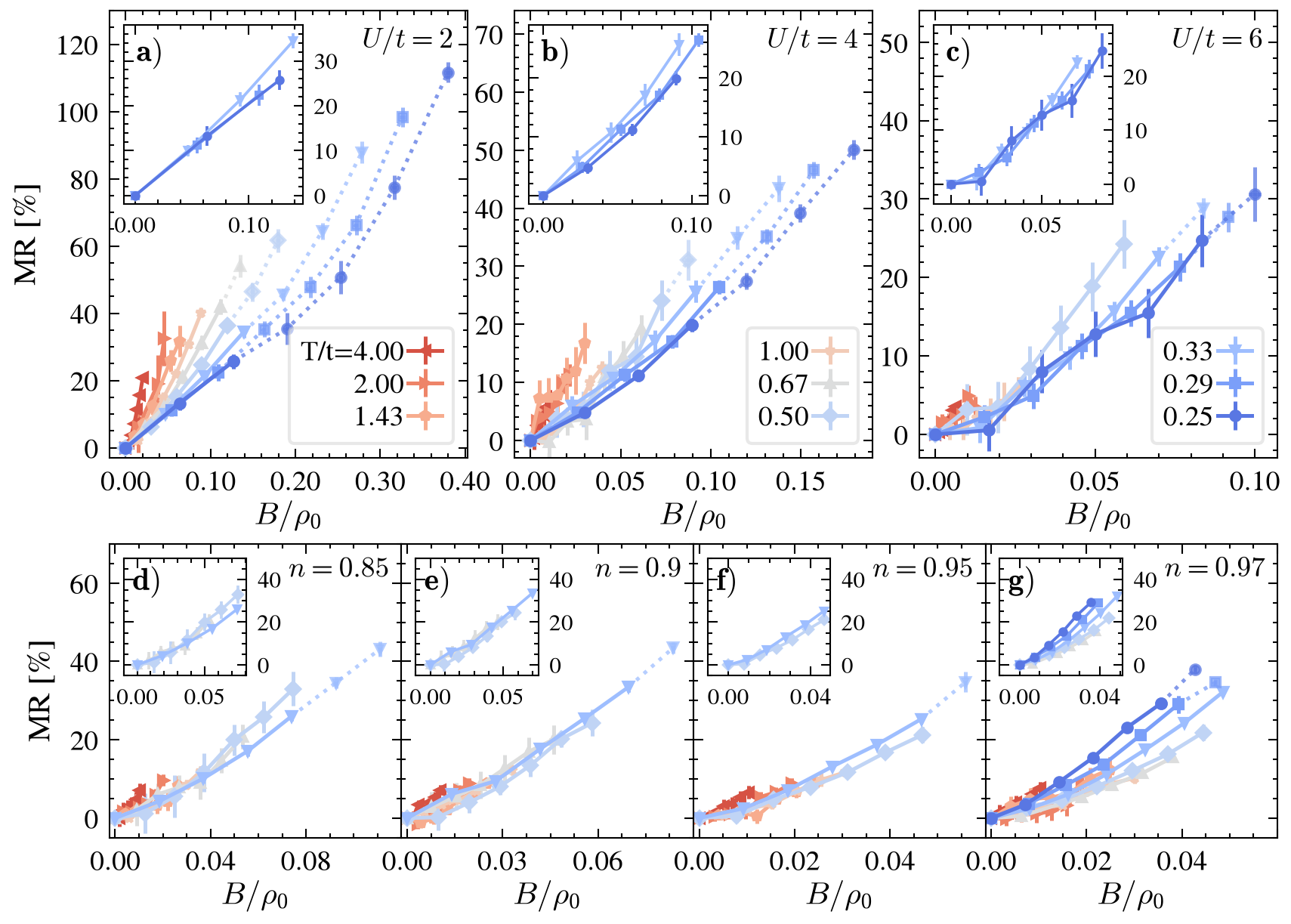}
    \caption{Kohler plots: magnetoresistance vs. the magnetic field strength $B$ divided by the zero-field resistivity $\rho_0$ for temperatures $T/t = 0.25-4.0$. (a--c) are interaction strengths $U/t = 2, 4, 6$ at fixed particle density $\langle n \rangle=0.8$. (d--g) are $\langle n\rangle = 0.85, 0.9, 0.95, 0.97$ at fixed $U/t = 4$. Solid and dotted lines denote points below and above the cutoff field $B^\star(U, T, n)$, respectively. Insets show zoomed views of the Kohler-obeying temperatures in each panel, restricted to below $B^\star$.}
\label{fig:kohler}
\end{figure*}

In Fig. \ref{fig:kohler}, we then show the MR vs. $B/\rho_0$ in Kohler plots.
For $U/t=2-6$ in panels (a)--(c), 
the MR curves at different temperatures do not generally overlap in our strange metallic parameter space, diverging dramatically at the highest temperatures and fields we access. When restricting to below the cutoff $B^\star$ and for our lowest temperatures $T/t=1/3, 1/3.5,$ and $1/4$, however, the curves remarkably collapse, and Kohler's rule is obeyed.
We observe similar trends in the doping dependence in panels (d)-(g), with $\langle n\rangle=0.85-0.97$; Kohler's again holds at low field and temperature, including up to $T/t = 1$ for the middle of our doping range. Strong violation for all $T$ and $B$ appears only at $\langle n\rangle = 0.97$ from proximity to the Mott insulator. As $\rho_0 \propto T$ with negligible residual resistivity, note that Kohler scaling here is equivalent to B/T scaling (see Supplemental Material Sec. VI).

\begin{figure}[!htbp]
    {\includegraphics[width=0.9 \columnwidth]{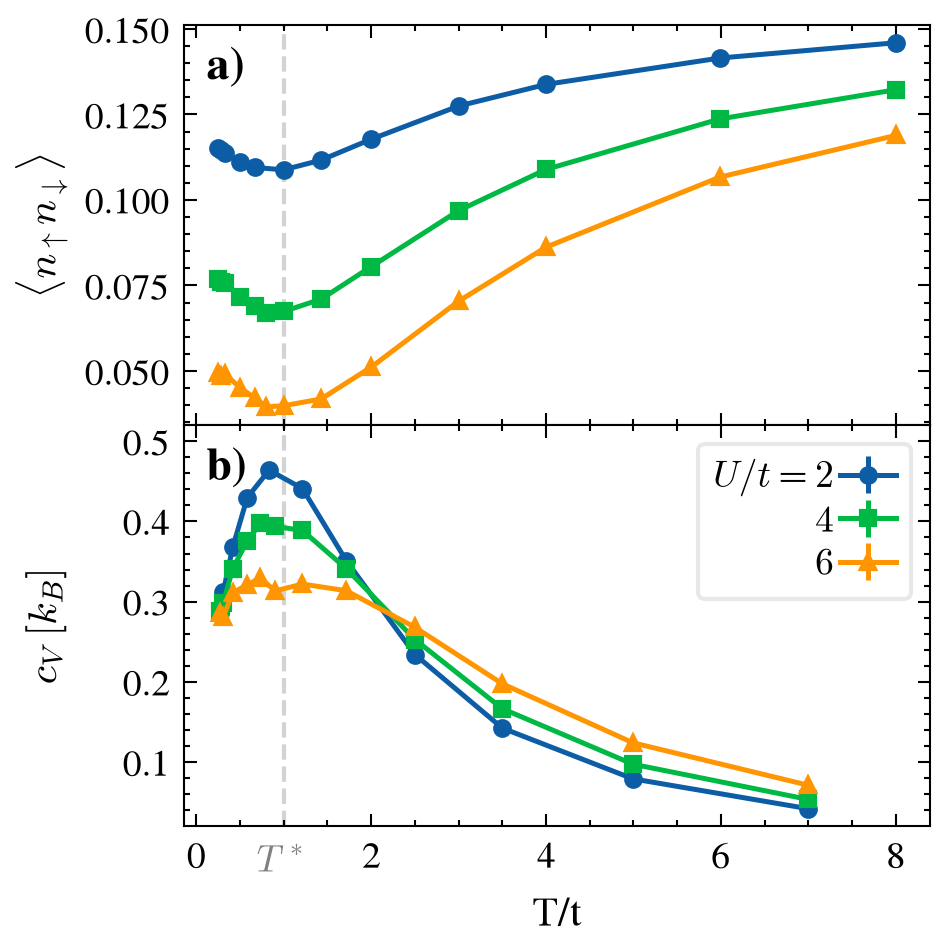}}
    \caption{Temperature dependence of the (a) double occupancy $\langle d\rangle = \langle n_\uparrow n_\downarrow \rangle$ and (b) specific heat $c_V$ for interaction strengths $U/t = 2, 4, 6$, and magnetic field strength $B = 1/64$. 
    Each curve exhibits a minimum/maximum at $T^\star \approx t$, separating a low-$T$ quantum coherent regime from a high-$T$ thermally disordered one.}
    \label{fig:doubleocc}
\end{figure}

Finally, we observe a temperature scale similar to where Kohler is obeyed in the thermodynamics. Fig. \ref{fig:doubleocc} shows the average per-site double occupancy $\langle d \rangle=\langle n_\uparrow n_\downarrow \rangle$ and specific heat $c_V$, both of which exhibit extrema at $T^\star/t \approx 1$. The upturn in $\langle d \rangle$ below this crossover temperature $T^\star$ signals the onset of local singlet states between sites \cite{paiva_signatures_2001}, while $c_V$ broadly peaks from the release of entropy as these short-range spin correlations develop. Kohler's rule holds in this regime. 

\textit{Discussion.---} 
Why is Kohler's rule, which itself makes no reference to temperature, obeyed primarily below $T/t \approx 1$? Comparing Figs. \ref{fig:rho}(b)-(d) and \ref{fig:kohler}(a)-(c), our coldest $U/t=6$ temperatures violate MIR yet still obey Kohler, implying that the reason is not quasiparticle coherence. 

From Fig. \ref{fig:doubleocc}, we argue that the answer is instead quantum coherence.
As the system cools below the temperature $T^\star/t \approx 1$ separately identified from the thermodynamics, the low energy, thermally accessible states become entangled, with singlet formation between sites and accompanying fluctuations in double occupancy \cite{paiva_signatures_2001}.
Both $\sigma_{xy}$ and the MR are thermal averages over these many-body states and can be expanded in flux-enclosing hopping loops on the lattice~\cite{zhang_hall_2026}.
$T^\star$ thus denotes a general crossover into a regime where quantum coherent states dominate the magnetotransport.

Kohler's rule in this quantum coherent regime appears highly robust, seeming to hold even for increasing $U$. In part by disrupting quasiparticle formation, strengthened correlations resist the influence of the magnetic field and raise the cutoff $B^\star$, stabilizing Kohler over a wider range of fields. 
This interpretation does not necessitate quasiparticles, nor for $B/\rho_0$ to be taken as a proxy to $\omega_c \tau$ in the conventional sense of cyclotron orbits.
Though the exact mechanism remains an open question,
the scaling collapse of MR with respect to $B/\rho_0$, lost only upon approach to half-filling, may be more fundamental than previously thought; strong correlations or anomalous resistivity do not themselves appear to break it, only the loss of metallicity altogether.

In strongly correlated materials, MR scaling is highly material specific -- perhaps this points to a hidden parameter dependence in our model not yet uncovered in this work. Investigating the MR in other parts of the Hubbard phase diagram may shed some light on this question. In particular, is Kohler still obeyed when particle-hole symmetry is broken by next-nearest neighbor hopping t', which dramatically affects $T^\star$ and qualitative behaviors in the Hall channel \cite{zhang_hall_2026}? While the sign problem precludes a full parameter sweep, extending this study to electron doping, where the Fermi surface is generally sharper and quasiparticles more coherent \cite{wang_probing_2025}, would offer a more feasible extension of this work.

\medskip
\textit{Acknowledgments.---}
We acknowledge helpful discussions with Nigel Hussey. This work was supported by the U.S. Department of Energy (DOE), Office of Basic Energy Sciences, Division of Materials Sciences and Engineering. Computational work was performed on the Sherlock cluster at Stanford University and on resources of the National Energy Research Scientific Computing Center, supported by the U.S. DOE under Contract no. DE-AC02-05CH11231. R. Jin was further supported by the National Science Foundation (NSF) Graduate Research Fellowship Program (GRFP) and the Stanford Graduate Fellowship (SGF). E. Z. Zhang was further supported by the Geballe Laboratory for Advanced Materials (GLAM) Postdoctoral Fellowship. 

\bibliography{ref}

\end{document}


\preprint{}

\title{Supplemental Material for ``Kohler's Rule Holds in the Strange Metallic Hubbard Model''}%

\author{Rebekah Jin}
\email{rebjin@stanford.edu}
\affiliation{Department of Applied Physics, Stanford University, CA 94305, USA}
\affiliation{Stanford Institute for Materials and Energy Sciences,
SLAC National Accelerator Laboratory, 2575 Sand Hill Road, Menlo Park, CA 94025, USA}

\author{Emily Z. Zhang}
\affiliation{Stanford Institute for Materials and Energy Sciences,
SLAC National Accelerator Laboratory, 2575 Sand Hill Road, Menlo Park, CA 94025, USA}
\affiliation{Geballe Laboratory for Advanced Materials, Stanford University, California 94305, USA}

\author{Jixun K. Ding}
\affiliation{School of Physics and Astronomy, Monash University, Clayton, VIC 3150, Australia}
\affiliation{Department of Applied Physics, Stanford University, CA 94305, USA}
\affiliation{Stanford Institute for Materials and Energy Sciences,
SLAC National Accelerator Laboratory, 2575 Sand Hill Road, Menlo Park, CA 94025, USA}

\author{Brian Moritz}
\affiliation{Stanford Institute for Materials and Energy Sciences,
SLAC National Accelerator Laboratory, 2575 Sand Hill Road, Menlo Park, CA 94025, USA}

\author{Thomas P. Devereaux}
\email{tpd@stanford.edu}
\affiliation{Stanford Institute for Materials and Energy Sciences,
SLAC National Accelerator Laboratory, 2575 Sand Hill Road, Menlo Park, CA 94025, USA}
\affiliation{
Department of Materials Science and Engineering, Stanford University, Stanford, CA 94305, USA}
\affiliation{Geballe Laboratory for Advanced Materials, Stanford University, California 94305, USA}

\date{\today}%

\maketitle

\onecolumngrid
\newpage

\section{Simulation parameters} 
\label{sup:sim_params}
DQMC data in this work are obtained from simulations performed using $4\times 10^3 - 4\times 10^4$ warm-up sweeps and $4 \times 10^5 - 4 \times 10^6$ measurement sweeps through the auxiliary field. We run $100-300$ independently seeded Markov chains for each set of parameters. For all parameter values, the imaginary time discretization interval $\Delta\tau \leq 0.1/t$, which is sufficient for negligible Trotter error, and the number of imaginary time slices $L = \beta/\Delta \tau \geq 10$. 

In all simulations, multiple equal-time measurements are taken in each full measurement sweep through the auxiliary field, while unequal-time measurements are taken every few full measurement sweeps. 
Each Markov chain with M measurement sweeps collects ML/5 equal-time measurements, and M/2 unequal-time measurements. 
The mean and standard error of equal-time observables (double occupancy, specific heat, density, etc.) are estimated via jackknife resampling of independent Markov chains. The chemical potential is iteratively tuned such that $|n_{\mathrm{target}}-\langle n\rangle| \leq 1 \times 10^{-4}$.

\section{Analytic continuation methods}
\label{sup:maxent}
We briefly outline the Maximum Entropy (MaxEnt) methods employed in this work. Further details are found in Refs. \cite{jarrell_bayesian_1996, reymbaut_maximum_2015, reymbaut_maximum_2017} as well as an upcoming manuscript.

\subsection{Electric current operator}
The electric current operator is $\mathbf{J} = (-e)\mathbf{J}_N$, where the particle current
$\mathbf{J}_N = \dot{\mathbf{P}} = (i/\hbar)[H, \mathbf{P}]$ is the rate of change of the
polarization operator $\mathbf{P} = \sum_i \mathbf{R}_i n_i$, the total position weighted by site occupation. For the Hubbard-Hofstadter Hamiltonian %
the total electric current operator then becomes
\begin{align}
    \label{current}
    \mathbf{J} &= \frac{i(-e)}{\hbar} t \sum_{ij\sigma} \exp\left[i\varphi_{ij}\right]
    (\mathbf{R}_i - \mathbf{R}_j) c^\dagger_{i\sigma} c_{j\sigma} \\
    &= \frac{ie}{2\hbar} t \sum_{ij\sigma} (\mathbf{R}_j - \mathbf{R}_i)
    \left[\exp[i\varphi_{ij}] c^\dagger_{i\sigma} c_{j\sigma}
    - \exp[i\varphi_{ji}] c^\dagger_{j\sigma} c_{i\sigma}\right],
\end{align}
with $t$ the hopping integral between nearest neighbor sites, $\phi_{ij}$ the Peierls phase, and $c_{i \sigma}^{\dagger}$/$c_{i \sigma}$ the creation/annihilation operator for an electron on site $i$ with spin $\sigma$. 

\subsection{Analytic continuation of optical conductivities}
In linear response, an applied electric field proportionally drives a current $\mathbf{J} = \hat{\sigma}\mathbf{E}$, where $\hat{\sigma}$ is the electric conductivity tensor.
$\hat{\sigma}$ can be expressed in terms of retarded current-current correlation functions through a Kubo formula
\begin{equation}
    \sigma_{\alpha \beta}(\omega) = \frac{i}{\omega} \chi^R_{\alpha \beta} (\omega)
\end{equation}
where
\begin{equation}
    \label{eq:retarded_chi}
    \chi^R_{\alpha\beta}(\omega) = \frac{-i}{\hbar V} \int_{-\infty}^{\infty} dt\,
    \theta(t) \left\langle \left[ J_\alpha(t), J_\beta(0) \right] \right\rangle_0
    e^{i\omega t}
\end{equation}
$J$ is the total electric current operator of Eq. \ref{current} which time-evolves according to the interaction representation (where the perturbation to the Hamiltonian by the external electric field is written $H' = -\textbf{J}\cdot \textbf{A}$ with $\textbf{E} = -\partial_t \textbf{A}$), $\langle \rangle_0$ is the unperturbed thermodynamic expectation value, and $\theta(t)$ is the step function.

DQMC simulations estimate the unequal imaginary time electric current-current correlator
\begin{equation}
    \chi_{\alpha\beta}(\tau) = \frac{1}{V}\langle J_{\alpha}(\tau) J_{\beta}(0)\rangle
\end{equation}
which must be analytically continued to the real frequency retarded correlator in Eq. \ref{eq:retarded_chi} to obtain the conductivity. 

\subsubsection{Longitudinal case}
    
A full derivation of the following may be found in \cite{mahan_many_particle_2000}. Here, we simply state for reference that for $\alpha=\beta$, $\chi(\tau)$ relates to $\chi^R(\omega)$ through
\begin{equation}
    \chi_{xx}(\tau) 
    = \int_{-\infty}^\infty d\omega K(\tau, \omega) \frac{-\mathrm{Im}[\chi^R_{xx}(\omega)]}{\pi \omega} 
    = \int_{-\infty}^\infty d\omega K(\tau, \omega) \frac{\mathrm{Re}[\sigma_{xx}(\omega)]}{\pi}
\end{equation}
where 
\begin{equation}
    K(\tau, \omega) = \frac{\omega e^{-\tau \omega}}{1-e^{-\beta \omega}}
\end{equation}
is known as the (Bosonic) kernel.
To invert the above integral and extract the optical conductivity, we apply the standard MaxEnt analytic continuation algorithm \cite{jarrell_bayesian_1996}, which selects the most probable spectral function by simultaneously optimizing goodness-of-fit while regularizing against a prior model function. For all spectra, we use flat model functions and select hyper-parameter $\alpha$ using the ``BT” method \cite{BT}. 

\subsubsection{Transverse case}
For $\alpha\neq\beta$, we have instead
\begin{equation}
    -i \chi_{xy}(\tau) 
    = \int d\omega K(\tau, \omega) \frac{\mathrm{Re}[\chi^R_{xy}(\omega)]}{\pi \omega} 
    = \int d\omega K(\tau, \omega) \frac{\mathrm{Im}[\sigma_{xy}(\omega)]}{\pi}
\end{equation}
Due to the definition of its regularizing entropy term, MaxEnt as-is cannot extract non-positive definite spectra like $\mathrm{Im}[\sigma_{xy}]$, which is antisymmetric in frequency. To circumvent this, we use the subtraction method described in \cite{reymbaut_maximum_2015, reymbaut_maximum_2017}, first analytically continuing the composite correlator
\begin{equation}
    \label{eq:composite}
    \lambda \chi_{xx}(\tau) -i \chi_{xy}(\tau)
    = \int d\omega K(\tau, \omega) \bigg[\lambda \frac{\mathrm{Re}[\sigma_{xx}(\omega)]}{\pi} + \frac{\mathrm{Im}[\sigma_{xy}(\omega)]}{\pi} \bigg]
\end{equation}
(using a constrained MaxEnt method described in the following section) before then subtracting off the separately continued $\mathrm{Re}[\sigma_{xx}]$ to obtain $\mathrm{Im}[\sigma_{xy}]$. As the off-diagonal conductivity is typically small compared to $\sigma_{xx}$, the summed spectral function is generally positive over all relevant frequencies for scaling parameter $\lambda = 1$. Where $\sigma_{xy}$ becomes comparatively large, as at the higher fields considered here, we increase $\lambda$ up to $10$ as needed to restore positivity.

Finally, we apply a Kramers-Kronig transform to obtain Re$[\sigma_{xy}]$:
\begin{equation}
    \mathrm{Re}\!\left[\sigma_{xy}(\omega)\right]
    = \frac{1}{\omega}\,\mathrm{p.v.}\!\!\int_{-\infty}^{\infty}
      \frac{\omega'\,\mathrm{Im}\!\left[\sigma_{xy}(\omega')\right]}{\omega'-\omega}
      \frac{d\omega'}{\pi}.
\end{equation}
The D.C. conductivities used to calculate the resistivity and MR are extracted by interpolating these real component spectra through $\omega=0$, as discussed in the main text. The mean and standard error of all MaxEnt analytic continuation results are estimated via bootstrap resampling of independent Markov chains using 200 bootstrap resamples, with the subtraction performed within each resample.

\subsection{Constrained MaxEnt}
For the composite analytic continuation described above, we use a newly developed constrained MaxEnt technique. We reserve a full description for an upcoming paper but provide a brief summary of the motivation and method here. 

Standard MaxEnt maximizes the following functional
\begin{equation}
    \label{eq:Q}
    Q[A] = \alpha S[A] - \chi^2[A]/2
\end{equation}
to determine the most likely spectral function $A^\star \in \mathbb{R}^N$, where $S$ is the regularizing entropy, $\chi^2$ is the goodness-of-fit, and $\alpha$ is a hyper-parameter which balances the relative priority of both (full definitions may be found in \cite{jarrell_bayesian_1996}). In this work, the optical conductivities we calculate are either even/odd in $\omega$, but there is nothing in standard MaxEnt which enforces physical symmetries; the optimization is completely unconstrained. This is not generally an issue for diagonal conductivities, which have both sufficient signal-to-noise to produce approximately symmetric spectra and may also be optimized over a reduced positive frequency space.

For the composite analytic continuation described by Eq. \ref{eq:composite}, however, the off-diagonal component Im$[\sigma_{xy}]$ has poor signal-to-noise, and the spectra returned by composite MaxEnt and subtraction are often severely asymmetric. 
The problem worsens with smaller Fermion sign at lower temperatures, making a reliable off-diagonal conductivity impossible to extract.

To address this issue, we maximize Eq. \ref{eq:Q} subject to explicit symmetry constraints. Writing $A_{tot} = A_{xx} + A_{xy}$,  we enforce the off-diagonal antisymmetry $A_{xy}(\omega) = -A_{xy}(-\omega)$ as a series of $N/2$ linear constraints $B A_{tot}=b$, where
\begin{align}
    B & = \begin{bmatrix} I_{N/2} & J_{N/2} \end{bmatrix}\\
    b &= 2 A_{xx}\big|_{\omega > 0}
\end{align}
where $J$ is the exchange matrix and $B \in \mathbb{R}^{N/2 \times N}$.
This encodes that $A_{tot}$ be the sum of a symmetric, separately continued $A_{xx}$ plus a yet undetermined antisymmetric modification $A_{xy}$. $\mathrm{Im}[\sigma_{xy}]$ then follows by subtraction as before.

\begin{figure*}[!htbp]
   \centering
    {\includegraphics[scale = 0.4]{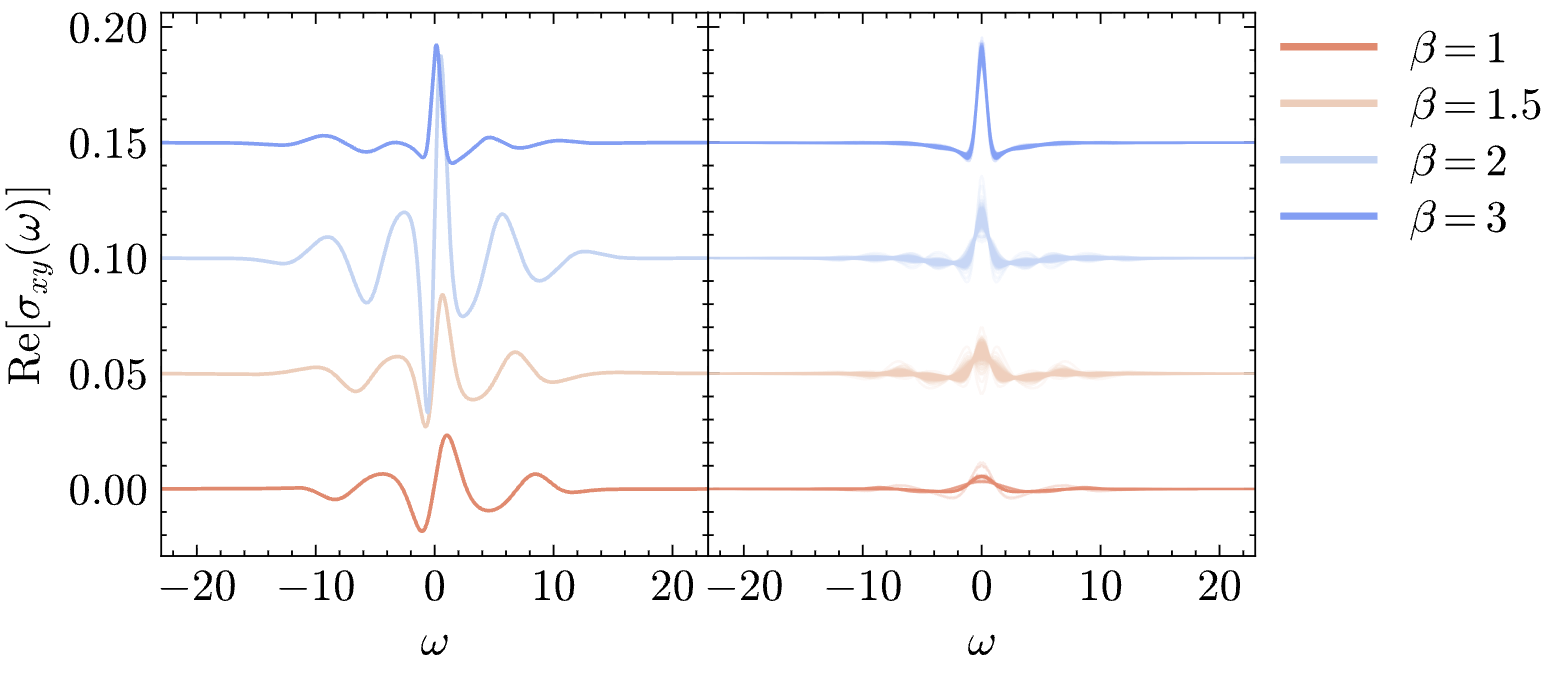}}
    \caption{$\mathrm{Re}[\sigma_{xy}(\omega)]$ from unconstrained (left) and constrained (right) MaxEnt at $U/t = 6$, $\langle n \rangle = 0.8$, and magnetic field strength $\Phi/\Phi_0 = 1$. Curves are offset vertically by $0.05$ for visibility. The left panel shows a single
    bootstrap sample; the right shows 200 bootstrap samples overlaid.}
    \label{fig:constrained}
\end{figure*}

We use CVXPY \cite{diamond2016cvxpy, agrawal2018rewriting}, a Python-embedded modeling language for convex optimization and the CLARABEL solver \cite{Clarabel_2024}. A comparison of standard and constrained MaxEnt for the off-diagonal conductivity is shown in Fig. \ref{fig:constrained}.

\subsection{Example optical conductivity spectra}
\begin{figure*}[!htbp]
   \centering
   {\includegraphics[scale = 0.8]{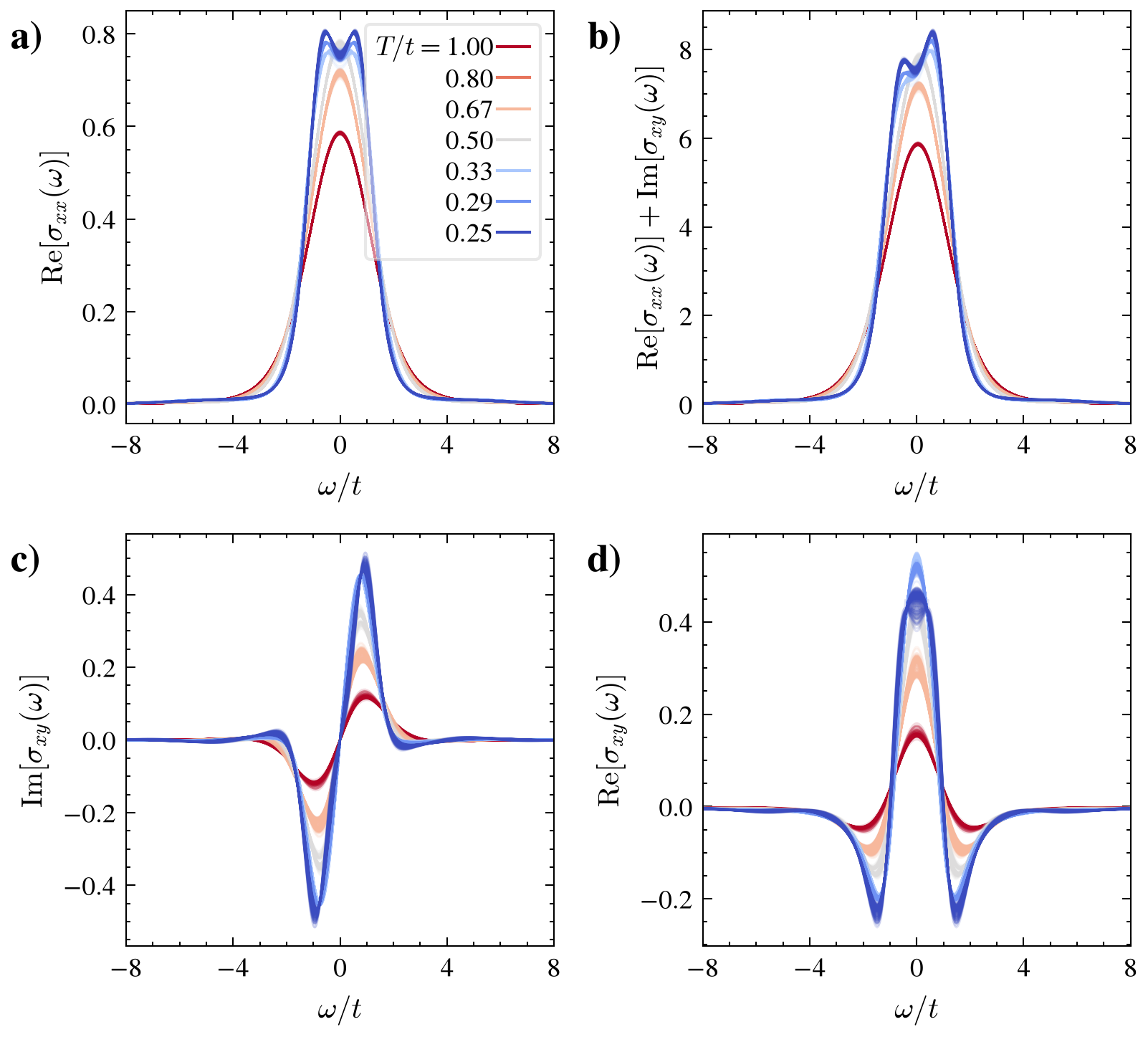}}
   \caption{Analytically continued optical conductivities a) $\mathrm{Re}[\sigma_{xx}(\omega)]$, b) $\lambda \mathrm{Re}[\sigma_{xx}(\omega)] + \mathrm{Im}[\sigma_{xy}(\omega)]$, c) $\mathrm{Im}[\sigma_{xy}(\omega)]$, and d) $\mathrm{Re}[\sigma_{xy}(\omega)]$ in units of $e^2/\hbar$ at $U/t=2$, particle density $\langle n \rangle = 0.8$, and magnetic field strength $\Phi/\Phi_0 = 8$.}
   \label{fig:optical_cond}
\end{figure*}

In fig. \ref{fig:optical_cond} we show representative $\mathrm{Re}[\sigma_{xx}(\omega)]$, $\lambda \mathrm{Re}[\sigma_{xx}(\omega)] + \mathrm{Im}[\sigma_{xy}(\omega)]$, $\mathrm{Im}[\sigma_{xy}(\omega)]$, and $\mathrm{Re}[\sigma_{xy}(\omega)]$, the frequency-resolved optical conductivity spectral functions obtained via the MaxEnt analytic continuation process outlined in the previous section. In this example with magnetic field strength $\Phi/\Phi_0 = 8$, $\sigma_{xy}$ becomes comparable to $\sigma_{xx}$, and we accordingly set scaling parameter $\lambda = 10$. We emphasize again that accurate resistivities at such high fields require a reliable off-diagonal conductivity, made newly possible by the constrained MaxEnt method.

\section{Finite size}
\begin{figure*}[!htbp]
   \centering
   {\includegraphics[scale = 0.65]{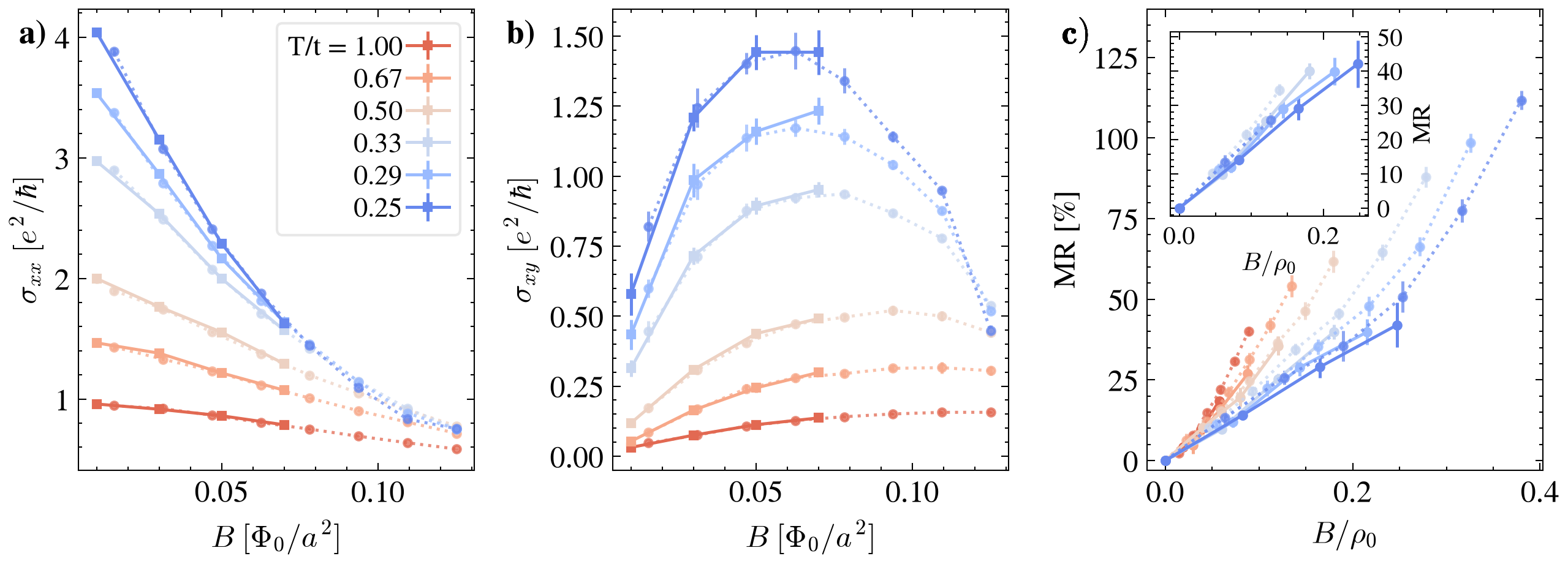}}
   \caption{Finite-size comparison at $U/t=2$ and particle density $\langle n \rangle = 0.8$. Dotted lines/circles and solid lines/squares denote an 8$\times$8 and 10$\times$10 lattice, respectively. (a) D.C. longitudinal conductivity $\sigma_{xx}$ and (b) transverse conductivity $\sigma_{xy}$ as a function of magnetic field $B$ (in units of flux quanta per plaquette). (c) Kohler plot, the magnetoresistance as a function of $B/\rho_0$, with inset showing a zoomed view of the low-field, low-temperature region.}
   \label{fig:finite_size}
\end{figure*}

To assess any possible finite-size effects, we repeat our calculations at $U/t=2$ on a larger $10\times10$ lattice and compare against the $8\times8$ results from the main text in Fig.~\ref{fig:finite_size}. Across the full range of temperatures and magnetic fields we simulate, both the longitudinal conductivity $\sigma_{xx}$ and transverse conductivity $\sigma_{xy}$ show close quantitative agreement between the two lattice sizes.

The Kohler plot shows close agreement as well. The MR as a function of $B/\rho_0$ on the $8\times8$ and $10\times10$ lattices coincides closely across our full field range. In particular, the $10\times10$ results exhibit the same B-linearity and low-temperature, low-field adherence to Kohler's rule as the $8\times8$. We conclude that our results on the validity of Kohler's rule do not suffer from a significant finite-size effect, and that our $8\times8$ results are representative of the thermodynamic limit for the interaction strengths studied here.

\section{Fermion sign}
\begin{figure*}[!htbp]
   \centering
   {\includegraphics[scale = 0.9]{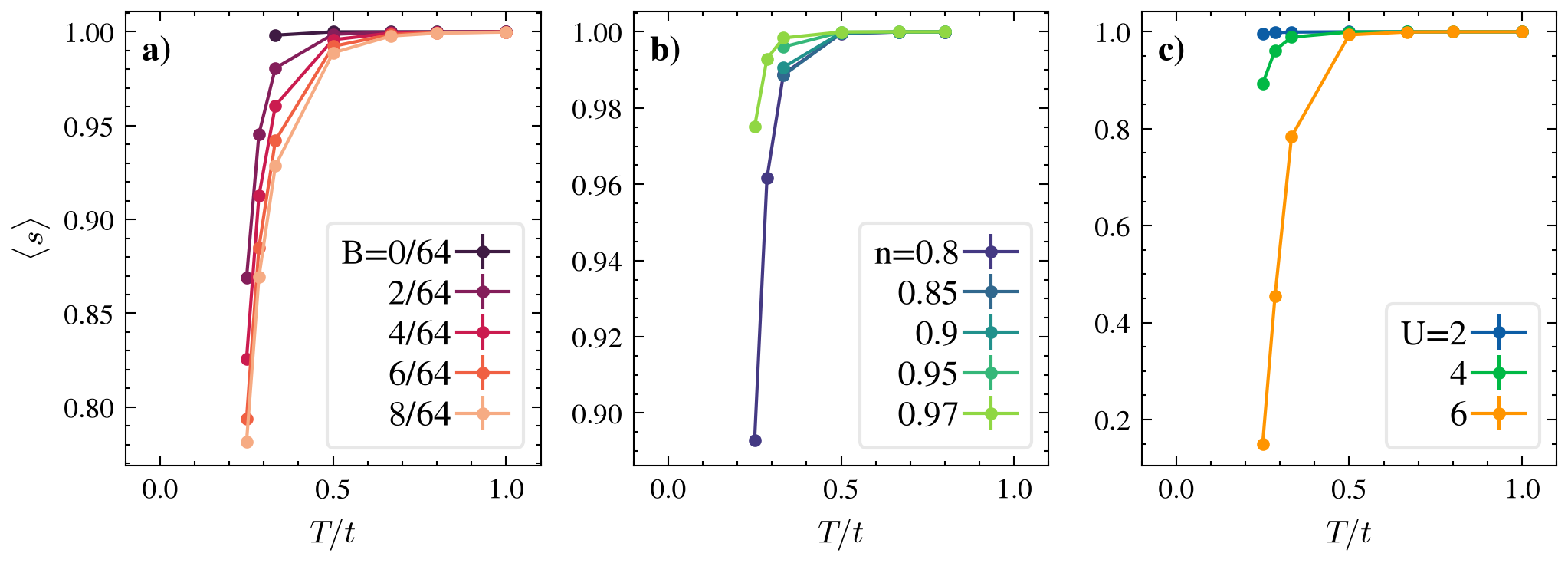}}
   \caption{Average fermion sign $\langle s \rangle$ vs. temperature for
    a) $U/t = 4$, $\langle n \rangle = 0.8$, $B = 0, 2, 4, 6, 8$ flux quanta
    (in units of $1/64$ per plaquette).
    b) $U/t = 4$, $B = 1/64$, $\langle n \rangle = 0.8, 0.85, 0.9, 0.95, 0.97$.
    c) $\langle n \rangle = 0.8$, $B = 1/64$, $U/t = 2, 4, 6$.
    Error bars are from jackknife resamples over measurement bins.}
   \label{fig:sign}
\end{figure*}

In Fig. \ref{fig:sign}, we show the Fermion sign dependence on our simulation parameters, which generally worsens for lower temperatures, stronger interaction strengths, and stronger fields. For now, the sign precludes a more thorough examination of the MR across the Hubbard phase diagram, which may be ameliorated with a greater volume of data; we leave this to future work.

\section{Doping dependence}
\label{sup:doping}
\begin{figure*}[!htbp]
   \centering
   {\includegraphics[scale = 0.95]{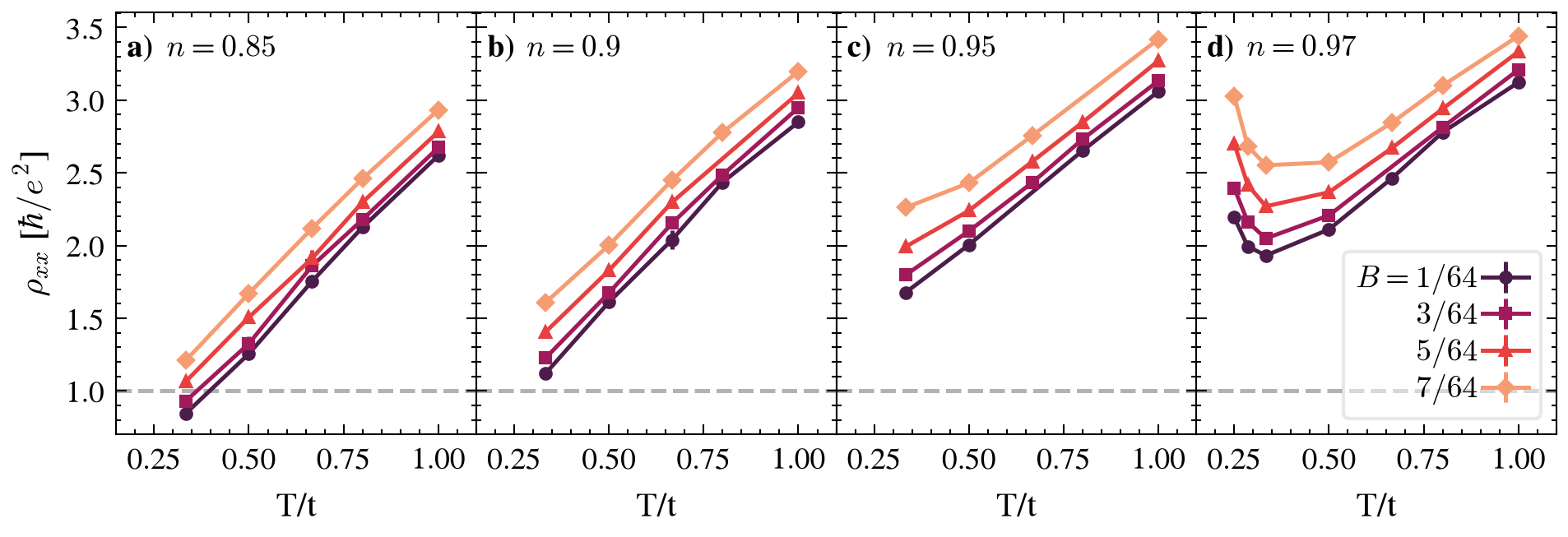}}
   \caption{Temperature and field dependence of the longitudinal resistivity $\rho_{xx}$ for $U/t=4$ and a) $n=0.85$, b) 0.9, c) 0.95, d) 0.97. The dashed horizontal line at $\hbar/e^2$ denotes the MIR limit.}
   \label{fig:rho_doping}
\end{figure*}

In this section, we show the full doping dependence of various measurements omitted from the main text for brevity. In Fig. \ref{fig:rho_doping} we show the longitudinal resistivity $\rho_{xx}$ for dopings $\langle n \rangle = 0.85, 0.9, 0.95, 0.97$. Fewer charge carriers closer to half-filling raises the resistivity, linear in $T$ except for our smallest hole-doping $\langle n \rangle =0.97$.

\begin{figure*}[!htbp]
   \centering
   {\includegraphics[scale = 0.95]{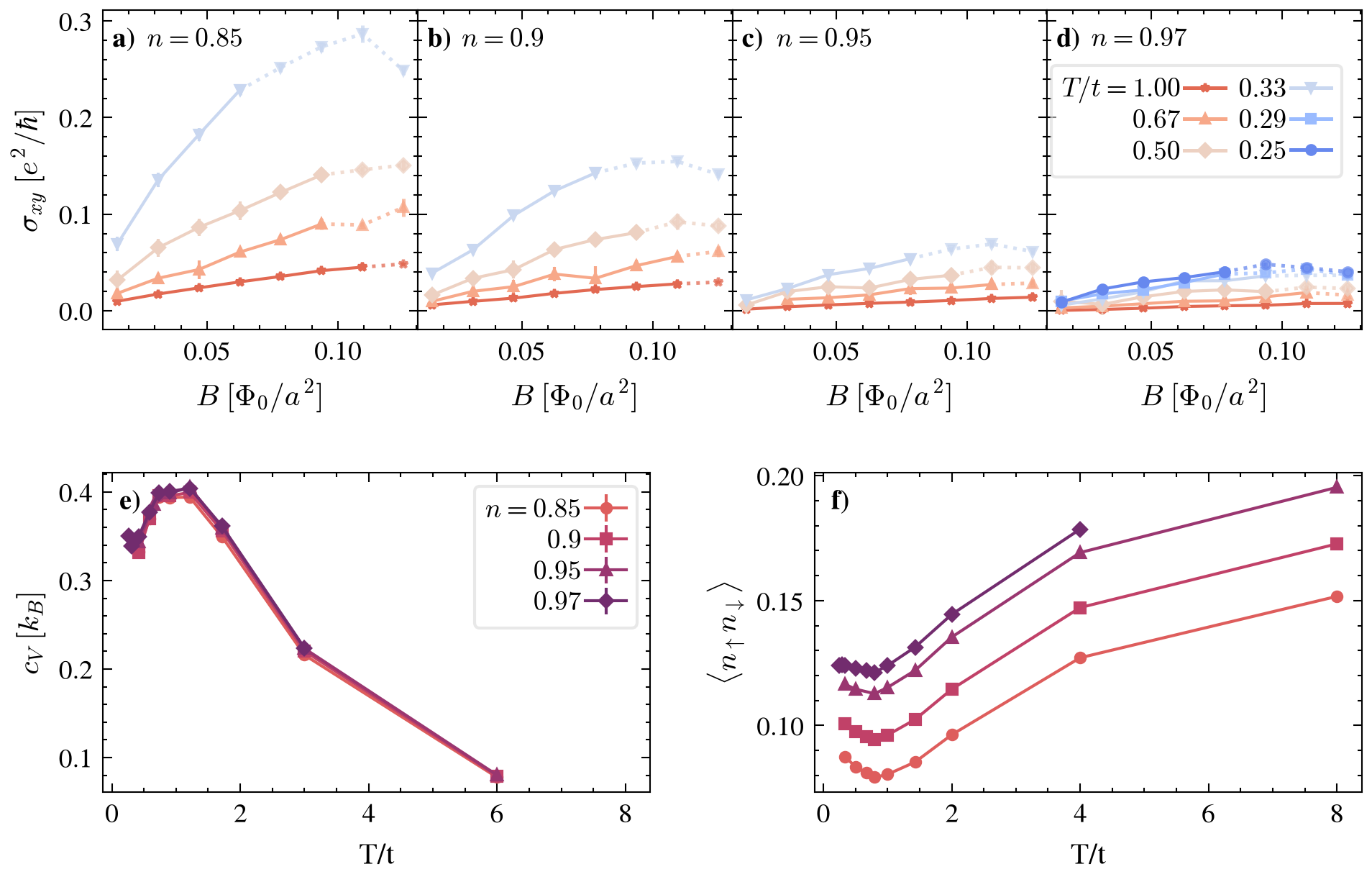}}
   \caption{a)--d) DC Hall conductivity $\sigma_{xy}(B)$ in units of $e^2/\hbar$ at $U/t=4$ for particle densities $\langle n \rangle = 0.85,\ 0.90,\ 0.95,$ and $0.97$, at temperatures $T/t = 1.00,\ 0.67,\ 0.50,\ 0.33,\ 0.29,$ and $0.25$ (the two lowest are available only at $\langle n \rangle = 0.97$). Solid lines indicate fields below the cutoff $B^\star$ beyond which $\sigma_{xy}$ is no longer reliable; dotted lines lie above it. e) Specific heat $c_V$ and f) double occupancy $\langle n_\uparrow n_\downarrow \rangle$ versus temperature for the same densities, at $U/t=4$ and $\Phi/\Phi_0 = 1$.}
   \label{fig:cutoffs}
\end{figure*}

In Fig. \ref{fig:cutoffs}, we show the full transverse conductivity $\sigma_{xy}$, specific heat $c_V$, and double occupancy $\langle d\rangle$ for the same dopings. The cutoff field $B^\star$ does not change as strongly with doping as with $U$, only slightly shifting out to higher field closer to half-filling. The crossover temperature $T^\star$ also does not appear to depend on the doping, remaining at an approximately constant $T^\star/t = 1$ for our entire density range.

\section{Alternative magnetoresistance scalings}
\label{sec:alt}

For completeness, we discuss some relevant alternative MR scalings in the literature. The first is the so-called modified Kohler's rule, where the low-field MR $\propto \tan^2(\theta_H)$ with Hall angle $\theta_H = \tan^{-1}(\sigma_{xy}/\sigma_{xx})$. This alternative scaling, reported in underdoped cuprates \cite{harris_violation_1995}, has primarily been used as evidence for Anderson's two-lifetime theory \cite{anderson_hall_1991}. In Fig. \ref{}, we do not observe the MR curves at different temperatures collapsing, at least with not nearly the same robustness across our doping and interaction parameters as with conventional Kohler's rule. While not an indictment of any associated theories, this suggests that this form of anomalous scaling may derive from more material-specific features.

Including higher fields, another form of the MR adds the temperature and field-dependences in quadrature, with $\rho(B, T) = \rho(0, 0) + \sqrt{(\alpha T)^2 + (\gamma B)^2}$ with $\alpha, \gamma$ fixed coefficients \cite{hayes_scaling_2016}. This expression captures the high/low field B- and T-quadraticity/linearity. Quadrature MR has been linked to proximity to a quantum critical point in a variety of strongly correlated materials, and far-reaching correlations found between $\alpha$ and $\gamma$, the high-T and high-B linear coefficients, suggest a common origin \cite{hayes_scaling_2016}. This is a particular form of the more general $B/T$ scaling, in which different temperatures of $(\rho(B, T)-\rho(0, 0))/T$ plotted vs. $B/T$ collapse \cite{berben_compartmentalizing_2022}. Note that for $\rho \propto T$ with $\rho(0, 0) = 0$, Kohler and $B/T$ scaling are indistinguishable. 
Key experimental observations of anomalous MR and its associations to quantum criticality report $T$-linear $\rho$ in the presence of a sizable residual resistivity, up to one third of the 300K value in Bi2201~\cite{kohler_Bi2201_2021}. As we do not observe a significant $\rho(0, 0)$ in our model, this remains an open question which we do not address in this work, especially as our MR analysis is primarily in the weak-field limit. 

\begin{figure*}[!htbp]
   \centering
   {\includegraphics[scale = 0.7]{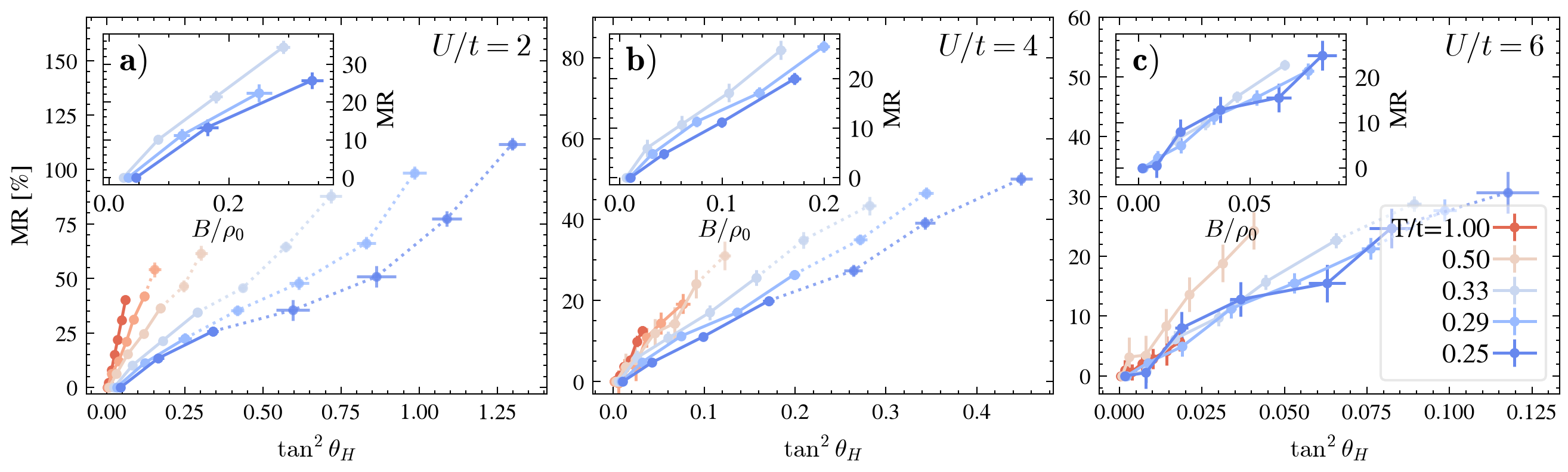}}
   \caption{Test of modified Kohler MR scaling. (a--c) MR vs.\ $\tan^2\theta_H$ (modified Kohler's rule) for 8x8, $n=0.8$, $U/t = 2, 4, 6$; insets show behavior below $B^*$ and $T^*$.}
   \label{fig:modified}
\end{figure*}

Finally, we mention here but do not plot an ``extended" Kohler's rule, which shows that some experimental Kohler violations originate from a hidden temperature-dependence of the carrier density $n(T)$ \cite{xu_extended_2021, luo_kohlers_2001}. The rule is inapplicable here; our chemical potential is tuned to a fixed particle density for each temperature simulation as described in \ref{sup:sim_params}.

\clearpage
\bibliography{ref}